\documentclass[12pt]{article}
\newcommand{\be}{\begin{equation}}
\newcommand{\ee}{ \end{equation}}

\newfont{\msbm}{msbm9 at 11pt}
\newfont{\msbms}{msbm5 at 8pt}
\newcommand{\C}{\mbox{\msbm C}}

\newfont{\smsbm}{msbm10 at 9pt}
\newfont{\gmsbm}{msbm10 at 18pt}
\newfont{\Gmsbm}{msbm10 at 24pt}

\input{tcilatex}

\begin{document}

\begin{center}
\textbf{\vspace{-2cm} \rightline{\mbox {\normalsize
{Lab/UFR-HEP/0112}}}} \vskip 1cm
 {\bf \Large Fractional
Statistics in terms of the $r$-Generalized Fibonacci  Sequences}

\vskip 2cm

\textbf{M. Rachidi$^{a}$ \footnote{\textit{D\'epartement de
Math\'ematiques et Informatique, Facult\'e des Sciences de Rabat,
B.P. 1014, Rabat, Morocco.}} , E.H. Saidi$^{a}$ and
J.Zerouaoui$^{a,b}$ \footnote{\textit{Laboratoire de Physique
Th\'eorique et Appliqu\'ee LPTA, Facult\'e des Sciences de
K\'enitra, Morocco }}}\\[0pt]
\

 (${a}$)  Facult\'e des Sciences, Lab/UFR-Physique des Hautes Energies,\\
Avenue Ibn Battouta B.P. 1014,Facult\'e des Sciences de Rabat,
Morocco\\[0pt]
\

 (${b}$) The Abdus Salam International Centre for Theoretical
 Physics,\\
34100, PO BOX586, 34100,Trieste, Italy.
\\[0pt]
\end{center}

\vskip 2cm

\begin{abstract}

 We develop the  basis of the two dimensional generalized quantum
statistical systems by using results on $r$-generalized Fibonacci
sequences. According to the spin value $s$ of the
2d-quasiparticles, we distinguish four classes of quantum
statistical systems indexed by $ s=0,1/2:mod(1)$, $s=1/M:mod(1)$,
$s=n/M:mod(1)$ and $0\leq s\leq 1:mod(1)$. For quantum gases of
quasiparticles with $s=1/M:mod(1)$, $M\geq 2,$, we show that the
statistical weights
 densities $\rho_M$ are given by the integer hierarchies of
Fibonacci sequences. This is a remarkable result which envelopes
naturally the Fermi and Bose statistics and may be thought of as
an alternative way to the Haldane interpolating statistical
method.
\end{abstract}
PACS number: 05.30.-d/11.10.-z/11.30.Ly.\newline 2000 MSC : 40A05,
40A25 \newpage
 \tableofcontents

\newpage

\section{Introduction}

Quantum elementary excitations carrying generalized values of the
spin, to which we shall refer here after as quasi-particles, are
involved in the description of many aspects of the critical
behaviours of physical systems in two and two plus one dimensions
[1,2]. In effective field models of fractional quantum Hall (FQH)
liquids for instance [3], the elementary excitations carry
fractional values of the spin and
fractional electric charge Q. In this kind of systems, the spin angle $%
\theta $ is a topological quantity which is generated by a Bohm
Aharonov effect and eventually by a Berry phase for systems with
non zero space curvature geometries. For FQH hierarchical states
of generic level n, typically described by a $U(1)^{n}$ CS gauge
theory with a matrix coupling K of integer entries $K_{ij}$
${1}\leq {i,j}\leq {n}$ and  integers $l_{i}$, the spin angle
$\theta\over 2\pi$ , proportional to $
s=\sum_{i,j=1}^{n}{l_{i}(K^{-1})_{ij}}l_{j}$, is not necessary
half integer. The spin s takes rational values and in general real
ones. \par From the mathematical point of view, the parameter s
may, roughly speaking, be thought of as the highest weight of the
$SO(2)$ representations. As such there is no restriction on it
and, a priori, can take any value. This parameter, together with
the scaling dimension $\delta$  were at the basis of the
spectacular developments made over the two last decades in 2d
critical systems
 [4,5,6].\par In $2d$ conformal field theory (CFT), objects
carrying generalized values of the spin are elegantly described by
world sheet (WS) vertex field operators type $V_{\alpha }(\phi
)=exp(i\alpha \phi )$; where $\phi $ describes one or a collection
of $2d$ free scalars and $\alpha $ is a parameter related to the
square root of the spin s of $V_{\alpha }(\phi )$ [7]. Quantum
states with general values of the spin are also involved in the
description of the continuous spectrum of the Liouville theory
[8], in the integrable
thermal deformation of minimal models[9] including the ${%
\mbox{\msbm Z}}_{N}$ Potts ones [10] and those related with
$U_{q}(sl_{2})$ quantum group representations[11] especially the
periodic ones [12].\par In statistical mechanics of many bodies,
quasi-particles carrying generalized spins have been subject to
some interest in the past. In [13]; these quasi-particles have
been interpreted as virtual particles interpolating between bosons
and fermions. This is explicitly seen on the Haldane statistical
weight $W(\alpha )$ giving the number of accessible states of a
gas of $N$ quasi-particles occupying a group of $G$ states:
\begin{equation}
W(\alpha )=\frac{[G-(\alpha -1)(N-1)]!}{N![G-\alpha N-(1-\alpha )]!}
\end{equation}
In this equation, $\alpha $ is an interpolating parameter lying
between zero and one describing respectively bosons and fermions.
Later on we will give the exact relation between $\alpha $ and the
spin $s$.\par Starting from this statistical weight formula, Wu
has studied in[14] the thermodynamic properties of gases of
quasi-particles with exotic spins and derived the basic equations
of the general form of the distribution and the grand partition
function. These equations are exactly solved for bosons, fermions
and for the so called semions. Some of the results of the Wu study
were rederived in [15,16] by using a completely different manner
based on an extension of the Pauli exclusion principle [17]. The
key idea of this method may be summarized as follows. Considering
a gas of quasi-particles of spin $ 1/M$ mod $1$, $M$ integer
greater than one $(M\geq 2)$ and assuming that the quantum
statistics of quasi-particles of spin $1/M$ are governed by a
generalization of the Pauli exclusion principle according to which
no more than $(M-1)$ quasi-particles can live altogether on the
same quantum state. This principle extends the usual Pauli
principle for fermions, associated here to the special case $M=2$.
In [15,16], see also [18], it has been shown that one can solve
exactly several statistical mechanics quantities for this kind of
quantum gases. The obtained results are in complete agreement with
the Wu analysis [14].\par In this paper we reconsider the study of
[15,16] by examining the basis of such kind of quantum statistical
systems using results on combinatorial aspects of generalized
r-Fibonacci sequences. Instead of the interpolation formula of
Haldane eq(2), we propose a new statistical weight $\rho (N+G,M)$,
given by the following combinatorial expression of $M$-generalized
Fibonacci sequences:
\begin{equation}
\rho (n,M)=\sum_{g_{0}+2g_{1}+\cdots +Mg_{M-1}=n-M}\frac{(g_{0}+g_{1}+\cdots
+g_{M-1})!}{g_{0}!g_{1}!\cdots g_{M-1}!}p_{0}^{g_{0}}p_{1}^{g_{1}}\cdots
p_{M-1}^{g_{M-1}},
\end{equation}
where the $p_{i}$'s, $0\leq p_{i}\leq 1,$ are interpreted as
occupation probabilities. To have an idea on how this relation
applies in practice, it is instructive to consider the pure case
where the $p_{i}$'s are set to one and so the above formula
reduces to the well known M-generalized Fibonacci numbers namely:
\begin{equation}
\rho (n,M)=\sum_{g_{0}+2g_{1}+\cdots +Mg_{M-1}=n-M}\frac{(g_{0}+g_{1}+\cdots
+g_{M-1})!}{g_{0}!g_{1}!\cdots g_{M-1}!}.
\end{equation}
Moreover setting $g_{0}+g_{1}+\cdots +g_{M-1}=G$ or equivalently
$g_{1}+2g_{2}+\cdots +(M-1)g_{M-1}=N$, one obtains the generalized
weight density of a gas of quasi-particles of spin $1/M,$ whose
leading case $M=2$, gives just the usual fermions statistical
weight $W(1)=\rho (N+G,2)_{|g_{0}+g_{1}=G}={\frac{G!}{N!(G-N)!}}$.
\par
This paper is organized as follows. In section 2, we give the set
up of quantum statistics of quasi-particles carrying general
values of the spin and fix terminologies. In section 3, we review
some general features of the Haldane Statistics and compare it to
our approach. In section 4, we make an explicit study of the
quantum gas of quasi-particles of spin $s=1/3$ either by using
standard techniques or in terms of $M=3$-generalized Fibonacci
numbers. In section 5 we review the M-generalized Fibonacci
sequences and exhibit their connection with fractional statistics.
In section 6 we make a discussion and give our conclusion.

\section{Generalized QSM}

\indent Statistics is a quantum mechanical feature of identical
particles which governs the quantum behaviour of a collection of
$N$ particles in the large limit of $N$. Usually one distinguishes
two kinds of quantum statistical systems: bosons carrying integer
values of the spin $(s=0~mod1)$ and fermions carrying half odd
integers spins $(s=1/2~mod1)$. The thermodynamic statistical
properties of bosons and fermions are described by two different
quantum statistical distributions namely the Bose and Fermi
densities. In this section we want to discuss the general set up
of the quantum statistics of systems of quasi-particles carrying
general values of the spin. Then we give some results regarding
the quantum statistics of quasi-particles of spin $1/M $, $M \geq
2$.\par To start recall that in two dimensions, there is no
restriction on the values the spin s; it can take any value; i.e:
$0\leq s\leq 1$ mod $1$. Thus, according to the values of the spin
$s$, the quasi-particles describing the quantum excitations in 2d
systems may be classified in different kinds and so it is natural
to conjecture the existence of various types of quantum
statistical systems in two dimensions. These are given by the
following classification:

\begin{itemize}
\item  QSM; quantum statistical systems with quasi-particles carrying spins $%
\ s=0,1/2~(mod1)$.

\item  FQSM; quantum statistical systems with quasi-particles carrying
fractional values of $s$; i.e:$s=1/M:mod(1)$, where $\ M\geq 2$

\item  RQSM; quantum statistical systems with quasi-particles carrying spins
$s$ type: $s=n/M:$ $mod(1)$, where $M\geq 2$ and $M\neq k.n$ with k an
integer.

\item  GQSM; quantum statistical systems with quasi-particles carrying spins
$s$, with $0\leq s\leq 1:mod(1)$.
\end{itemize}

This classification, according to the values of the spin $s$, obeys
naturally to the following embedding:
\begin{equation}
QSM\subset FQSM\subset RQSM\subset GQSM.
\end{equation}
It is not difficult to see that QSM, describing bosons ( $s=0$ mod (1)) and
fermions ( $s=1/2$ mod (1)), is recovered as the boundary limit of FQSM by
taking respectively $M=+\infty $ and $M=2$. Similarly, FQSM may be viewed as
a particular case of RQSM by setting $n=1$. Finally, putting $s=n/M$ in GQSM
by restricting to quasi-particles with rational values of the spin, one
obtains RQSM.

From the above classification, we expect that several features of QSM may be
extended iteratively to the other statistical systems. For example the well
known fermionic number operator $N=[(-)^{F}]$ extends naturally as
\begin{eqnarray}
&<&\psi \mid N\mid \psi >=<\psi \mid expi(2\pi F)\mid \psi > \\
&=&expi(2\pi s),
\end{eqnarray}
where $\mid \psi >$ is the wave function of a given quasi-particle state of
spin s. The eigenvalues of the operator F are effectively given by the spin
of the $\mid \psi >$ eigenfuntions; namely $s=0:mod(1)$ for bosons, $%
s=1/2:mod(1)$ for fermions, $s=1/M:mod(1)$ for FQSM, $s=n/M:mod(1)$ for RQSM
and $0\leq s\leq 1:mod(1)$ for GQSM.

An other example of features of QSM which extends naturally to the
other quantum systems of eq(4) is the Pauli principle. This
exclusion principle may be generalized to all other quantum
statistical systems. This is easily seen by computing the trace of
fermionic number operator $expi(2\pi F)$ on representation spaces
involving all possible values of the spin; in particular for the
supermultiplets built of states ($s=0,$ $s=1/2)$ for QSM and their
generalizations ($s=j/M,:0\leq j\leq M-1$) for FQSM.

\begin{eqnarray}
Tr:expi(2\pi F) &=&1-1=0,\quad \mbox{for}\quad QSM, \\
Tr:expi(2\pi F) &=&\sum_{j=0}^{M-1}[expi({\frac{2\pi j}{{\ M}}})]=0,\quad %
\mbox{for}\quad FQSM,
\end{eqnarray}

Such relation extends to RQSM and GSSM as well. For the supersymmetric
multiplets ($s=jn/M,:0\leq j\leq M-1)$ of the RQSM systems; we have \
\begin{equation}
Tr:expi(2\pi F)=\sum_{j=0}^{M-1}[expi({\frac{2\pi jn}{{M}}})]=0,\quad %
\mbox{for}\quad RQSM,
\end{equation}
while for the multiplets ($s=\beta ,:0\leq \beta \leq 1)$ of GQSM systems,
we have the following formula:
\begin{equation}
Tr:expi(2\pi F)=\int_{o}^{1}ds[expi(2\pi s)]=0,\quad \mbox{ for }\quad GQSM.
\end{equation}

Having classified the various kinds of quantum statistical
systems, we turn now to describe briefly methods used in studying
gases of quasi-particles. There are several ways of doing. The
first way, due to Haldane, is based on the combinatorial formula
eq(1) which turns out to be the simplest and the natural weight
one may conjecture. This approach uses linear interpolating
relations between the Bosons and the Fermions and concerns
quasiparticles of spin $s$ lying between $1/2$ and $1$. Haldane
quasiparticles obey a generalized exclusion Pauli principle
although a clear statement of this principle for the continuous
values of $s$ is still lacking. A second approach which was
suggested in [19], follows the same line of reasoning as of
Haldane; but instead of looking for interpolating statistical
weight, one considers interpolating wave functions. In other words
if we denote by $\phi $ and $\psi $ respectively the Bose and
Fermi wave functions of a system of N particles, then the wave
function $\chi $ of gases of quasiparticles of spin $s$ is,
roughly speaking, given by:
\begin{equation}
\chi =\phi cos(2\pi s)+\psi sin(2\pi s);0\leq s\leq 1/2.
\end{equation}
However, a detailed analysis of this eq shows that for the case of
quasi-particles of spin $s=1/M$ one discovers exactly the
quasi-determinant of [15,16] describing the wave function of the
FQSM gas of spin $s=1/M$ \ (mod $1); $ with $M$ $\geq 2$.

A third way, which was developed concerns gases of quasiparticles
of spin $s=1/M$ (mod $1)$, is based on the idea of imposing from
the beginning the generalized Pauli exclusion principle according
to which no more than $(M-1)$ quasiparticles of spin $1/M$; $M\geq
2$ can live altogether on the same quantum state. In this
approach, one obtains explicit relations for FQSM extending the
Bose and Fermi analogue; in particular the following quantum
distributions
\begin{equation}
<n_{k}>=\frac{\xi }{1-\xi }-\frac{M\xi ^{M}}{1-\xi ^{M}},
\end{equation}
where we have set $\xi =e^{-\beta (\epsilon _{k}-\mu )}$and $M\geq 2$. This
eq offers an alternative to the standard quantum q-uon one given by $<n(q)>=%
\frac{\xi }{1-q\xi };$where $-1\leq q\leq 1$ and where for $q=1,-1,0$ we get
respectively the Bose, Fermi and Boltzman distributions. .

In the next section, we review the main lines of the Haldane
statistics eq(1).Then, we develop further the alternative we
considered in [15,16]. Using group theoretical methods and
combinatorial analysis we show that the rigorous results we
obtained are not fortuitous but are due to the existence of a link
between FQSM systems and a well known subject in mathematical
physics namely Fibonacci sequences hierarchies. To that purpose,
we shall proceed as follows:\par (a) first review the basis of the
Haldane approach; this method is based on a linear interpolating
formula between bosons and fermions. The simplicity of this way is
due to the fact that the weight density is due to the remarkable
properties of the special function $\beta (x,y)$. Non linear
interpolations between bosons and fermions were also considered in
[19]; they involve hypergeometric functions
$\mathcal{F(}a,b;\alpha ,\beta \mathcal{)}$ $\mathrm{and}$.\par
(b) Develop a new setting using rigorous analysis involving group
theoretical techniques and combinatorial formulas and show that
the obtained results are indeed linked to generalized Fibonacci
sequences.

\section{Haldane Statistics}

We start this section by describing some general features of the
Haldane statistics. Although exact solutions are still missing,
this statistics exhibits many remarkable features which make it
very special. Haldane statistics based on eq(1), corresponds in
fact to a linear interpolating statistics between bosons and
fermions. This linearity is a useful mathematical feature which
make the interpolating formula natural and simple to handle.
Indeed, setting
\begin{equation}
\begin{array}{ll}
x=x(\alpha )=N &  \\
y=y(\alpha )=G-\alpha N+(\alpha -1), &
\end{array}
\end{equation}
in eq(1), one sees that it reduces to the inverse of the well known di-gamma
special function $\beta (x,y)$; namely:
\begin{eqnarray}
W(x,y) &=&\frac{(x+y)!}{x!y!}, \\
W(x,y).\beta (x,y) &=&1.
\end{eqnarray}
Note that $y(\alpha )$ is a linear interpolation between the Bose point $%
y_{B}$ and the Fermi one $y_{F}$ respectively given by:
\begin{equation}
\begin{array}{ll}
y_{B}=y(1)=G-N &  \\
y_{F}=y(0)=G-1. &
\end{array}
\end{equation}
The interpolation eq(13) as well as the statistics of the boundary points
eq(16), shows that the Haldane interpolating parameter $\alpha $ and the
spin $s$ are related in the large $N$ limit as
\begin{equation}
\alpha =\frac{N(1-2s)}{N-2}\sim (1-2s),0\leq s\leq \frac{1}{2}.
\end{equation}
For $s=0$, and $s=\frac{1}{2}$, the parameter $\alpha $ is equal to one or
zero respectively. For the special case $\alpha =\frac{1}{2}$, corresponding
to semions of spin $s=\frac{1}{4}$, the Haldane statistics can be solved
exactly; one finds the following quantum distribution for semions is given
by:
\begin{equation}
<n_{j}(\alpha =1/2)>=\frac{1}{[\frac{1}{4}+exp2\beta (\epsilon _{j}-\mu
)]^{1/2}}
\end{equation}
For general values of $\alpha $, the situation is however complicated
because one has to solve non linear eqs of the type:
\begin{eqnarray}
<n_{j}> &=&
\frac{1}{w(e^{\beta(\epsilon_{j}-\mu)})+\alpha},\nonumber\\ \xi&=&
w(\xi)^{\alpha}{[1+w(\xi)]^{1-\alpha}}.
\end{eqnarray}
where $\xi= e^{\beta(\epsilon-\mu)}$, $\epsilon$ is the energy and
$\mu$ is the chemical potential.
\section{Gas of Quasiparticles of Spin $1/3$}

The gas of identical quasi-particles of spin $1/3$ is the simplest system
coming after the quantum Fermi gas. This is a quantum system of FQSM, which
may be studied by using two routes:

(i) using the Haldane interpolating statistical weight $W(\alpha)$
eq(1) and fixing the parameter as $\alpha=\frac{N}{3(N-2)} $; i.e
$s=1/3.$

(ii) solving the constraint equations imposed by FQSM according to which no
more than $(M-1)$ quasi-particles of spin $s=1/M$ can live altogether on the
same quantum state.

\subsection{Interpolating Statistics}

Following Haldane and using eq(17), the statistical weight $W_{s=1/3}$ of a
gas of $N$ quasi-particles of $s=1/3$ can be shown to be given by:
\begin{equation}
W_{s=1/3}={\frac{\Gamma \lbrack \frac{3(N-2)G-(N-1)(2N-6)}{3(N-2)}]\Gamma (N)%
}{\Gamma \lbrack \frac{3(N-2)G-2N^{2}-2N+6}{3(N-2)}]}}.
\end{equation}

This eq may be rewritten into an equivalent form as:
\begin{equation}
W_{s=1/3}=W(N,y),
\end{equation}
where now $y=\frac{1}{3(N-2)}[3(N-2)G-N^{2}+6-2N]$. Having these relations
at hand one can derive the thermodynamic statistical mechanical features of
this gas of quasi-particles. For example, the mean occupation number $%
<n_{k}> $ is obtained by solving eqs(19). For the case of a gas
with pure occupation probabilities, the partition function $z(\xi
)$ reads, at the statistical equilibrium, as:
\begin{equation}
z(\xi )=1+\xi +\xi ^{2}.
\end{equation}
Note that eq(22) lies between with the partition
function$z_{Fermi}(\xi )=1+\xi ,$ and $\ z_{Bose}(\xi
)=1+\sum_{n=1}^{\infty }{\xi }^{n}$ of individual fermions and
bosons of quantum Fermi and Bose gases respectively.
$z_{Fermi}(\xi )$ is given by just the sum of two terms as a
consequence of the Pauli exclusion principle contrary to the
bosonic gas where no exclusion rule exist. Therefore, the
polynomial eq(22) of order 2 reflects clearly that quasiparticles
of spin $s=1/3$ obeys a generalization exclusion statistics
principle according to which no more than two particles of spin
$s=1/3$ can live together on the same quantum state of the gas.
This selection rule is the simplest exclusion statistics coming
after the Pauli exclusion formula obeyed by fermions and
constitutes the leading term of a more general hierarchical
formula namely:
\begin{equation}
z(\xi )=1+\xi +...+\xi ^{M-1}.
\end{equation}
This relation extends eq(22) for the case of quasi-particles of
spins $\frac{1}{M},$ $M\geq 2$. In what follows we will show that
such relation may be also derived by using an other alternative
using combinatorial analysis.

\subsection{Generalized Exclusion Statistics}

We start by examining the special exclusion statistics obeyed by
quasi-particles with spin $s=1/3$ by using group theoretical methods and
combinatorial analysis. Then we work out the properties of the formula of
the statistical weight at equilibrium one gets, in particular its relation
to 3-generalized Fibonacci numbers. To do so, consider a system of $N$
quasi-particles of spin $s=1/M$ with $M=p+1$, occupying $G$ quantum states.
On each state of the $G$ ones, there can live altogether at most $p$
quasi-particles as required by the generalized exclusion statistics for
particles with spin $s=1/(p+1)$. To get the total number of accessible state
of the $p=2$ system, we proceed as follows: First divide the set of the
allowed $G$ quantum states into three kinds of sub-sets $g_{0},g_{1}$ and $%
g_{2}$ consisting respectively of $B_{0},B_{1}$ and $B_{2}$ boxes which can
contain zero, one and two quasi-particles of spin $s=1/3$ of the system with
occupation probabilities $p_{0}$, $p_{1}$ and $p_{2}$ with $\ 0\leq
p_{i}\leq 1$. Thus, we have the following equalities:
\begin{eqnarray}
g_{0}+g_{1}+g_{2} &=&G, \\
g_{1}+2g_{2} &=&N.
\end{eqnarray}
Since the $B_{0},B_{1}$ and $B_{2}$ boxes are indistinguishable, it follows
that for a given configuration of the quantum system; the number of
accessible states is no longer given by $%
G!p_{0}^{g_{0}}p_{1}^{g_{1}}p_{2}^{g_{2}}$ but rather by:
\begin{equation}
\frac{G!}{g_{0}!g_{1}!g_{2}!}p_{0}^{g_{0}}p_{1}^{g_{1}}p_{2}^{g_{2}}.
\end{equation}
For configurations associated with pure occupation probabilities;
i.e $p_{0}=p_{1}=p_{2}=1$, the above relation reduces to the well
known $\frac{G! }{g_{0}!g_{1}!g_{2}!}$ factor. Taking into account
all possible partitions of the number $N$ , the total number
$D_{3}(N,G)$ of accessible states of the quantum gas occupying G
quantum states reads then as:
\begin{equation}
D_{3}(N,G)=\sum_{g_{1},g_{2}\geq 0,g_{1}+2g_{2}=N}{\ \frac{G!}{%
g_{1}!g_{2}!(G-g_{1}-g_{2})!}}p_{0}^{g_{0}}p_{1}^{g_{1}}p_{2}^{g_{2}},
\end{equation}
where we have used eq(25). Furthermore solving $g_{1}$ in terms of $N$ and $%
g_{2}$ by help of eq(26), one may rewrite this equation into following
equivalent form.
\begin{equation}
D_{3}(N,G)=\sum_{g=0}^{[N/2]}\frac{G!}{g!(N-2g)!(G+g-N)!}%
p_{0}^{g_{0}}p_{1}^{g_{1}}p_{2}^{g_{2}},
\end{equation}
where we have set $g=g_{2}$ and where $[N/2]$ is the integer part of $N/2$.
Note in passing that this eq has some remarkable features; one of which is
that $D_{3}(G,N)$ contains the fermion statistical density $D_{2}(G,N)$ as
the leading term of the sum over the possible partitions of the $B_{2}$
boxes. In other words; $D_{3}(N,G)$ may be written, for pure probabilities,
as
\begin{equation}
D_{3}(N,G)=\frac{G!}{N!(G-N)!}+\sum_{g=1}^{[N/2]}\frac{G!}{g!(G-2g)!(G+g-N)!}%
,
\end{equation}
where the first term of the right hand side (rhs) of is just the statistical
weight $D_{2}(N,G)$ of $N$ fermions occupying $G$ states. The remaining $%
([N/2]-1)$ extra terms of eq(30) describe then a deviation above the $%
D_{2}(G,N)$ density and so it may be interpreted as the effects of
quasi-particles of spin $s=1/3$. Put differently, we have
\begin{equation}
D_{3}(G,N)=D_{2}(G,N)+\triangle _{32}(G,N),
\end{equation}
where $\triangle _{32}=D_{3}(G,N)-D_{2}(G,N)$, given by the second term of
the rhs of eq(32), describes the manifestation of FQSM.

\section{Fibonacci Sequences Setting}

A careful observation of the previous combinatorial analysis shows
that it has much to do with the combinatorial realization of the
generalized Fibonacci sequences[20], see also[21,22]. To exhibit
this connection explicitly, let us first give some elements on the
combinatorial aspects of these sequences, in particular for the
Fibonacci sequences of order M=2 and 3. To start let $
a_{0},a_{1}$ be some fixed real numbers with $a_{1}\not=0$ and
consider the sequence $\{V_{n}\}_{n=0}^{+\infty }$ defined by
$V_{n}=\rho (n+1,2)$ with,
\begin{equation}
\rho (n,2)=\sum_{g_{0}+2g_{1}=n-2}\frac{(g_{0}+g_{1})!}{g_{0}!g_{1}!}%
a_{0}^{g_{0}}a_{1}^{g_{1}},,::\mbox{ for every }::n\geq 3,
\end{equation}
and $\rho (2,2)=1$, $\rho (n,2)=0$ for $n=0,:1$. Using this eq it is not
difficult to verify that $V_{0}=0$, $V_{1}=1$ and
\begin{equation}
V_{n+1}=a_{0}V_{n}+a_{1}V_{n-1},\mbox{for every}:n\geq 1,
\end{equation}
implying in turns that $\{V_{n}\}_{n=0}^{+\infty }$ is nothing else a
weighted Fibonacci sequence with initial conditions specified by the numbers
$V_{0},V_{1}$ and weights $a_{0},a_{1}$. Note that for $a_{0}=a_{1}=1$, we
recover the classical Fibonacci numbers,
\begin{equation}
\rho (n,2)=\sum_{g_{0}+2g_{1}=n-2}\frac{(g_{0}+g_{1})!}{g_{0}!g_{1}!},,::%
\mbox{
for every }::n\geq 3.
\end{equation}
Observe also that up on imposing the extra constraint $g_{0}+g_{1}=G$, the $%
V_{n}$ is intimately related to the fermions number density $\mathcal{D}%
_{2}(G,N)$. Indeed putting the condition $g_{0}+g_{1}=G$ back into eq(36),
one discovers that sum over $g_{1}$ and $g_{0}$ with $g_{0}+2g_{1}=n-2$ is
completely fixed and so $n=N+G+2$ as well as:
\begin{equation}
\mathcal{D}_{2}(G,N)=\rho (N+G+2,2)|_{g_{0}+g_{1}=G},
\end{equation}
showing that the fermion density $\mathcal{D}_{2}(G,N)$ is given by a
restricted Fibonacci numbers sequence of rank two. An inspection of this
result shows that it is not a fortuitous feature and may be used to describe
not only fermions but also statistical weights of quantum gases of
fractional spin $s=1/M$. This observation is supported by several arguments,
above all the fact that Fibonacci sequences admit integer hierarchies in
perfect coherence with FQSM systems. To illustrate the idea, we consider in
what follows the $M=3$ case; i.e the case of a quantum gas of
quasi-particles of spin $1/3$. We begin however by giving some generalities
on 3-generalized Fibonacci sequences, then we derive the relation linking $%
D_{3}$ and $\rho (n,3)$. Later on we give the general result. Let $%
a_{0},a_{1},a_{2}$ be some fixed real numbers with $a_{2}\not=0$. \ If\ $%
a_{2}=0$ $but$ $a_{1}\not=0,$ one falls in the previous case. A
3-generalized Fibonacci sequences $\{V_{n}\}_{n=0}^{+\infty }$ is
roughly speaking defined by:
\begin{equation}
V_{n+1}=a_{0}V_{n}+a_{1}V_{n-1}+a_{2}V_{n-2},\mbox{for every}:n\geq 2,
\end{equation}
where $V_{0},V_{1},V_{2}$ and $a_{0},a_{1},a_{2}$ are respectively the
initial conditions and the weights of $\{V_{n}\}_{n=0}^{+\infty }$. Eq(38)
extends the usual Fibonacci sequences $\{V_{n}\}_{n=0}^{+\infty }$ obeying
eq(44), and which may also baptized as the rank 2 Fibonacci sequences; the
leading element of the Fibonacci hierarchy. . Recall by the way that
r-generalized Fibonacci sequences, which have been studied by using various
method and techniques, go beyond eqs(44) and (47), and play a fundamental
role in many fields of applied sciences and engineering [ ]. In practice
there are two ways to handle these sequences: the combinatorial realization
which we consider hereafter and the Binet representation. For the third
order of the hierarchy we are discussing, the combinatorial realization is
generated by:
\begin{equation}
\rho (n,3)=\sum_{g_{0}+2g_{1}+3g_{2}=n-3}\frac{(g_{0}+g_{1}+g_{2})!}{%
g_{0}!g_{1}!g_{2}!}a_{0}^{g_{0}}a_{1}^{g_{1}}a_{2}^{g_{2}},,::%
\mbox{ for
every }::n\geq 3,
\end{equation}
where we have set $\rho (3,3)=1$ and $\rho (j,3)=0$ for $0\leq j\leq 2$. To
check that the above eq is indeed a 3-generalized Fibonacci sequence, it is
enough to take $V_{n}=\rho (n+1,3)$. Straightforward calculations show that
we have $V_{0}=V_{1}=0$ and $V_{2}=1$ as well as:
\[
V_{n+1}=a_{0}V_{n}+a_{1}V_{n-1}+a_{2}V_{n-2},\mbox{for every}:n\geq 2.
\]
Note in passing that for $a_{0}=a_{1}=a_{2}=1$, the sequence $%
\{V_{n}\}_{n=0}^{+\infty }$ gives the well known 3-generalized Fibonacci
numbers, very familiar in the mathematical literature[M],[K]. In this case,
eq (45) becomes:
\begin{equation}
\rho (n,3)=\sum_{g_{0}+2g_{1}+3g_{2}=n-3}\frac{(g_{0}+g_{1}+g_{2})!}{%
g_{0}!g_{1}!g_{2}!},::\mbox{ for every }::n\geq 3,
\end{equation}
with $\rho (3,3)=1$ and $\rho (j,3)=0$ for $0\leq j\leq 2$. The beauty of
this relation is that it is linked to the density of the quantum gas of
quasi-particles of spin $s=1/3$ we studied earlier. Indeed, if instead of
the two constraint eqs (17-18), we consider rather the following relaxed
condition
\begin{equation}
g_{0}+2g_{1}+3g_{2}=G+N,
\end{equation}
obtained by adding eq(17) and eq(18), then the density $D_{3}(G,N)$, given
by eqs(28-29), is translated exactly to eq(46). Put differently $D_{3}(G,N)$
can be derived from the 3-generalized Fibonacci numbers by imposing either $%
g_{1}+2g_{2}=N$ or $g_{0}+g_{1}+g_{2}=G$. In the formulation based
on the Fibonacci sequences, the number of accessible states eq(19)
is now replaced by:
\begin{equation}
\frac{(g_{0}+g_{1}+g_{2})!}{g_{0}!g_{1}!g_{2}!}%
p_{0}^{g_{0}}p_{1}^{g_{1}}p_{2}^{g_{2}}.
\end{equation}
Taking into account of all possible configurations of the number $N+G$
partitions in eq(41), the new total number $\rho _{3}(N,G)=\rho (G+N-3,3)$
of accessible states is:
\begin{equation}
\rho _{3}(N,G)=\sum_{g_{0}+2g_{1}+3g_{2}=G+N}\frac{(g_{0}+g_{1}+g_{2})!}{%
g_{0}!g_{1}!g_{2}!}p_{0}^{g_{0}}p_{1}^{g_{1}}p_{2}^{g_{2}},
\end{equation}
which sould be compared with eq(39).Introducing the Dirac delta function $%
\delta (x)$, we can write,
\begin{equation}
D_{3}(G,N)=\delta (g_{0}+g_{1}+g_{2}-G)\rho (G+N-3,3).
\end{equation}
Substituting the delta function and solving $g_{1}$ in terms of $N$ and $%
g_{2}$, we can rewrite eq(20) into the following equivalent form:
\begin{equation}
D_{3}(N,G)=\sum_{g=0}^{[N/2]}\frac{G!}{g!(N-2g)!(G+g-N)!},
\end{equation}
where we have set $g=g_{2}$ and where $[N/2]$ is the integer part
of $N/2$. Having this relation at hand, we can write down all
thermodynamics properties of the gas of quasi-particles at
equilibrium.\newline

\section{More on FQSM}

In this section we want to extend some of the previous results to
quantum systems of $N$ quasi-particles of spin $1/M$; $M\geq 2$
using the combinatorial M-generalized Fibonacci sequences. We
start by generalizing the useful properties on the combinatorial
and the Binet expressions for the M-generalized Fibonacci
sequences.\par
Let $a_{0},a_{1},...,a_{M-1}$ be $(M-1)$ arbitrary real numbers with $%
a_{M-1}\not=$ 0 and $M\geq 2$. An M-generalized Fibonacci sequence $%
\{V_{n}\}_{n=0}^{+\infty }$ is defined as follows:
\begin{equation}
V_{n+1}=a_{0}V_{n}+a_{1}V_{n-1}+...+a_{M-1}V_{n-M+1},\mbox{for}:n\geq M-1
\end{equation}
where $V_{0},V_{1},...,V_{M-1}$ and $a_{0},a_{1},...,a_{M-1}$ are the
initial conditions the sequence. In case where $a_{i}=1$ for all i's and $%
V_{0}=V_{1}=...=V_{M-2}=0$ and $V_{M-1}=1$, we recover the M-generalized
Fibonacci numbers while for $a_{M-1}=0$ and $a_{M-2}\neq 0,$ one gets .$M-1$%
-generalized Fibonacci sequences to which one can apply the same analysis.
The combinatorial expression of $\{V_{n}\}_{n=0}^{+\infty }$, for $n\geq M$,
is realized as:
\begin{equation}
\rho (n,M)=\sum_{g_{0}+2g_{1}+\cdots +Mg_{M-1}=n-M}\frac{(g_{0}+g_{1}+\cdots
+g_{M-1})!}{g_{0}!g_{1}!\cdots g_{M-1}!}a_{0}^{g_{0}}a_{1}^{g_{1}}\cdots
a_{M-1}^{g_{M-1}},
\end{equation}
where we set $\rho (M,M)=1$ and $\rho (n,M)=0$ for $0\leq n\leq M-1$.
Starting from the following identity

\[
\frac{(k_0+\cdots +k_{M-1})!}{k_0!k_1!\cdots k_{M-1}!}= \sum_{j=0}^{M-1}%
\frac{(k_0+\cdots k_{M-1}-1)!} {k_0!\cdots k_{j-1}!(k_j-1)!k_{j+1}\cdots
k_{M-1}!}
\]

\noindent we derive easily that,

\[
\rho (n+1,M)=a_{0}\rho (n,M)+a_{1}\rho (n-1,M)+\cdots +a_{M-1}\rho
(n-M+1,M),
\]
Thus the sequence $\{W_{n}\}_{n=0}^{+\infty }$ defined by $W_{n}=\rho
(n+1,M) $, is a sequence whose initial conditions are such that $%
W_{0}=W_{1}=...=W_{M-2}=0$ and $W_{M-1}=1$. More generally, the
combinatorial expression of the sequence (46) can be given, for every $n\geq
M$, in terms of $\rho (n,M)$ $(47)$ as follows.
\begin{equation}
V_{n}=A_{0}\rho (n,M)+A_{1}\rho (n-1,M)+\cdots +A_{M-1}\rho (n-M+1),
\end{equation}
where
\[
A_{m}=a_{M-1}V_{m}+a_{M-2}V_{m+1}+\cdots +a_{m}V_{M-1};\mbox{
for }::m=0,1,\cdots ,M-1,
\]
and where $\rho (M,M)=1$ and $\rho (j,M)=0$ for $j<M$.

In the Binet representation of $M$-generalized Fibonacci sequence, the $%
V_{n}^{\prime }s$ are expressed as polynomials in the eigenvalues of:
\begin{equation}
P(X)=X^{M}-a_{0}X^{M-1}-\cdots -a_{M-2}X-a_{M-1},
\end{equation}

The point is that to every sequence $\{V_{n}\}_{n=0}^{+\infty }$ of rank $M,$
we associate the polynomial $P(X)$ above

whose roots $\lambda _{0},\lambda _{1},\cdots ,\lambda _{s}$ have
respectively the multiplicities $m_{0},m_{1},\cdots ,m_{s}$. The Binet
realisation of the general term $V_{n}$ of the sequence reads in terms of
the $\lambda _{i}^{\prime }s$ as:
\begin{equation}
V_{n}=\displaystyle\sum_{l=1}^{s}\;\displaystyle\sum_{j=0}^{m_{s}-1}\beta
_{l,j}n^{j}\lambda _{l}^{n},
\end{equation}
where the $\beta _{l,j}^{\prime }s$ are obtained from initial conditions by
solving the following system of $M$ linear equations
\begin{equation}
\displaystyle\sum_{l=1}^{s}\;\displaystyle\sum_{j=0}^{m_{s}-1}\beta
_{l,j}n^{j}\lambda _{l}^{n}=V_{n},:\mbox{ for }:n=0,1,...,M-1.
\end{equation}
Such expressions are known in the mathematical literature as the Binet
formula of $\{V_{n}\}_{n\geq 0}$. For the leading n=2 case where the $%
\lambda _{j}$ solutions of the characteristic polynomial $%
P(x)=x^{2}-a_{0}x-a_{1}$ have are simple to handle, we have
\begin{eqnarray}
V_{n} &=&\sum_{j=1}^{2}B_{j}\lambda _{j}^{n},\mbox{ for
        }\lambda _{1}\not=\lambda _{2}, \\
V_{n} &=&(B_{1}+B_{2}n)\lambda _{1}^{n},\mbox{ for
        }\lambda _{1}=\lambda _{2},
\end{eqnarray}
where the $B_{j}$'s are related to $\alpha _{0}$ and $\alpha _{1}$ as
\begin{eqnarray}
\sum_{j=1}^{2}B_{j}\lambda _{j}^{n} &=&\alpha _{n};\,\,\,\,\,\,\,n=0,1(%
\mbox{
for }\lambda _{1}\not=\lambda _{2}) \\
(B_{1}+B_{2}n)\lambda _{1}^{n} &=&\alpha _{n},\,\,\,\,\,\,\,n=0,1(%
\mbox{ for
}\lambda _{1}=\lambda _{2}).
\end{eqnarray}

Such eqs may written in a more explicit way by specifying the
explicit form of the $\lambda _{j}^{\prime }s$ in terms of the
initial parameters.

Having given some tools on Fibonacci sequences hierarchies, let us turn now
to expose briefly the link with FQSM systems. To that purpose, we reconsider
the gas described in section 4 but now the quasi-particles have spin $s=1/M$
with $M=P+1$, $P\geq 1$. In this case eqs(27-28) extend as
\begin{eqnarray}
g_{0}+g_{1}+g_{2}+....+g_{p} &=&G \\
g_{1}+2g_{2}+3g_{3}+....+Pg_{P} &=&N
\end{eqnarray}
The density of accessible states of the gas is,
\begin{equation}
\mathcal{D}_{P+1}(N,G)=\sum_{
\begin{array}{l}
{g_{1},g_{2},....,g_{P}\geq 0} \\
\sum_{j=1}^{P}j.g_{j}=N
\end{array}
}G![(G-\sum_{k=1}^{P}G_{P})!\prod_{l=1}^{P}(g_{l}!)]^{-1}
\end{equation}

Eq(37) exhibits several features, some of them extends properties of the $%
\mathcal{D}_{3}(N,G)$ statistical weight discussed earlier. This concerns
for instance the interpretation of $\mathcal{D}_{P+1}(N,G)$ as a deviation
above the $\mathcal{D}_{P}(N,G)$ weight of quasi-particles of spin $1/P$.
Indeed rewriting eq(54) by singling out the sum over $g_{P}$; i.e,
\begin{equation}
\mathcal{D}_{P+1}(N,G)=\sum_{g_{P}=0}^{[N/P]}\frac{1}{g_{P}!}\sum_{
\begin{array}{l}
{g_{1},g_{2},....,g_{p-1}\geq 0} \\
\sum_{j=1}^{P-1}j.g_{j}=N-Pg_{P}
\end{array}
}\frac{G!}{g_{1}!g_{2}!....,g_{P-1}!(G-\sum_{j=1}^{P}g_{j})!}
\end{equation}
one sees that the leading term in the sum over $g_{P}$; ie for $g_{P}=0$, is
just the weight density $\mathcal{D}_{P}(N,G)$ of a gas of $N$
quasi-particles of spin $1/P$ $($mod $P)$. Put differently, eq(55) may
usually decomposed as
\begin{equation}
\mathcal{D}_{P+1}(N,G)=\mathcal{D}_{P}(N,G)+\triangle _{P+1,P};
\end{equation}
where $\triangle _{P+1,P}$ measures the derivation from $\mathcal{D}%
_{P+1}(N,G)$ to $\mathcal{D}_{P}(N,G)$ densities and reads as
\begin{equation}
\triangle _{P+1,P}=\sum_{g_{P}=1}^{[N/P]}\frac{1}{g_{P}!}\sum_{
\begin{array}{l}
{g_{1},g_{2},....,g_{P-1}\geq 0} \\
\sum_{j=1}^{P-1}j.g_{j}=N-Pg_{P}
\end{array}
}\frac{G!}{g_{1}!g_{2}!....,g_{P-1}!(G-\sum_{j=1}^{P}g_{j})!}
\end{equation}
This eq is just a generalization of eqs(32-33); it is valid for
any value of $P$ greater than three and obeys quite remarkable
properties inherited from the generalized Fibonacci combinatorial
method we have been using. Eq(57) may be also thought as a
manifestation of tne exclusion statistics in FQSM stating that no
more than $P$ quasi-particles of spin $\frac{1}{P+1}$ can live
altogether on the same quantum state. \ It satisfies equally the
following splitting
\begin{equation}
\mathcal{D}_{p+1}(N,G)=[\mathcal{D}_{2}(N,G)+\sum_{j=3}^{p-1}\mathcal{D}%
_{j}(N,G)]+\sum_{j=2}^{p}\triangle _{j+1,j};
\end{equation}
in agreement with the embedding (4). Here the $\triangle _{j+1,j}$'s, given
by similar relations to (57), are deviations from $D_{j+1}$ to $D_{j}$
densities.

\section{Conclusion}

In this paper we have developed a new way to approach generalized quantum
statistical mechanics of gases of quasi-particles. Our method, which is
based on group theoretical analysis and combinatorial representation of
M-generalized Fibonacci sequences, can be viewed as an alternative to the
Haldane approach. To do so, we have first reviewed the setting of
generalized quantum statistics which we have classified into four classes
according to the values of the spin s of the quasi-particles: (1) the usual
QSM for $s=0,1/2 \: mod(1)$, (2) FQSM for $s=1/M \: mod(1)$, (3) RQSM for $%
s=n/M \: mod(1)$ and (4) GQSM for $0\leq s \leq 1: mod(1)$. This
classification obeys the inclusions: $QSM \subset FQSM \subset RQSM \subset
GQSM $.

Then we have reviewed the interpolating analysis of Haldane based
on eq(1) and showed that it is equivalent to the method we
developed in [14,15] using a generalization of the Pauli exclusion
rule as a general principle. \vskip 0.2cm Next motivated by a set
of observations; in particular the embedding eq(4), we have
developed an other way to deal with FQSM gases. Our method is
based on the following weight density:
\begin{equation}
\rho(n,M)=\sum_{g_0+2g_1+\cdots +Mg_{M-1}=n-M} \frac{(g_0+g_1+\cdots
+g_{M-1})!}{g_0!g_1!\cdots g_{M-1}!} p_0^{g_0}p_1^{g_1}\cdots
p_{M-1}^{g_{M-1}},
\end{equation}
where the $p_i$'s are dilution probabilities. Up on imposing
eqs(24-25) or equivalently eq(38), we rederive the known results.
We believe that our way of doing offer a powerful manner to deal
with such quantum systems. Our motivations for this claim are:

(1)$\:$ Eq(60) is obtained by using a rigorous analysis based on a
group theoretical method and a combinatorial analysis.

(2)$\: $Eq(60) is in agreement with the embedding $QSM \subset
FQSM $. It recovers standard results of QSM and describes
naturally any kind of quantum gas of quasi particles of spin
$s=1/M,\: \: M \geq 2 $..

(3)$\:$ Eq(60) is linked to a well known subject in mathematical
physics; namely the combinatorial aspect of the M-generalized
Fibonacci numbers. In this setting QSM is described by the
standard Fibonacci number sequences while FQSM is described by its
integer hierarchies. QSM appears then as the leading term of the
generalized Fibonacci hierarchy. In this context we expect that
RQSM and GQSM would be also described by means of similar
analysis.

(4)$\:$ Finally eq(60) incorporates in a natural way the dilution
property of quantum gases carried by the $p_i$ parameters.
Therefore diluted quantum gases of FQSM are described by integer
hierarchies of Fibonacci sequences.

\vskip 3cm \textbf{Acknowledgment}\newline
One of the authors; J.Z. would like to thank the AS-ICTP and the High Energy
Section for hospitality. He would also to acknowledge the Student program at
ICTP for invitation.\newline

This research work has been supported by the program PARS number: 372-98
CNR. \newpage

\vskip 1cm \textbf{References}
\begin{enumerate}
\item[[1]]   Daniel Friedan, Zong-an Qiu, Stephen Shenker, Phys.Rev.Lett.52:1575-1578,1984 \
Phys.Lett.B151:37-43,1985 \\C.Itzykson, J.M Drouffe, Statistical
Field Theory (1989) Cambridge, UK: Univ. Pr (1989).\\
 F.Wilczek, Fractional Statistics
and Anyon supraconductivity, WS Singapore(1990).
\item[[2]]R.B Laughlin, Quantum Hall Effet, R.E Prange and S.M
Girvin (Eds).
\item[[3]] X.G. Wen, Y.S. Wu and Y. Hatsugai,
Nucl.Phys.B422:476-494,1994\\ Xiao-Gang Wen ,Topological orders
and Edge Excitations in FQH States,PRINT-95-148 (MIT).\\ A. Jellal
 Int.J.Theor.Phys.37:2751-2755,1998\\
 I. Benkaddour, A. El Rhalami, E.H. Saidi;
 Non Trivial extension of the (1+2)-Poincare algebra and conformal
 Invariance on the Boundary of $AdS_3$; hep-th/0007142,To appear in Euro Jour
 C (2001).
\item[[4]]  A.A. Belavin,A.M Polyakov and A.B
Zamolodchikov,Nucl.Phys.B241:333-380,1984.
\item[[5]]  A.B.Zamolodchikov,  Rev.Math.Phys.1:197-234,1990.Adv.Stud.Pure Math.19:641-674,1989
\item[[6]]  J.Cardy, Conformal Invariance and Universality in Finite
 size Scaling,J.PHYS. A17 (1984) L385-L387.
\item[[7]]  Oleg Andreev, Boris Feigin
Nucl.Phys.B433:685-711,1995,\\P. Di Vecchia, M. Frau, A. Lerda and
S.Sciuto Nucl.Phys.B298:527,1988.
\item[[8]]  Andre LeClair and Dennis Nemeschansky, Mod.Phys.Lett.A11:139-146,1996.\\
Denis Bernard and Andre Leclair, Nucl.Phys.B399:709-748,1993.
\item[[9]]A.B Zamolodchikov, Nucl.Phys.B366:122-134,1991 \\
G.Mussardo Phys Report 218 (1992)215.
\item[[10]]A.Kadiri, E. H.Saidi, M. B.Sedra and J.Zerouaoui, ICTP
preprint IC/94/216.
\item[[11]]A.Leclair and C.Vafa Nucl Phys B401 (1993)413
\item[[12]]E.H.Saidi, M.B.Sedra and J.Zerouaoui, Class. Quant.Grav. 12 (1995) 1576.
\newline
E.H.Saidi, M.B.Sedra and J.Zerouaoui, Class. Quant. Grav. 12
(1995) 2705.
\item[[13]]F. D. H. Haldane, Phys. Rev. Lett. 67 (1991) 37.
\item[[14]]Y. S. Wu, Phys. Rev. Lett. 73, N7 (1994).
\item[[15]]E. H. Saidi and J. Zerouaoui, Revue Marocaine des Sciences
Physique N2 (2000).
\item[[16]]M. Rachidi, E. H. Saidi and J. Zerouaoui, Phys. Lett.
B409 (1997) 394.
\item[[17]]E. H. Saidi, ICTP preprint IC/95/299.\newline
E. H. Saidi, Proceeding de la journee de Physique Statistique;
Avril (1995), FST Settat, Morocco.
\item[[18]]J. Zerouaoui, Doctorat d'Etat, Rabat Morocco (2000).
\item[[19]] N.Fahssi, Lab/UFR-PHE (2000)Unpublished.

\item[[20]]T. P. Dence, Ratios of generalized Fibonacci sequences.
Fibonacci Quartely 25 (1987), 137-143.
 F. Dubeau, On 2-generalized Fibonacci sequences. Fibonacci Quartely 27 (1989),
221-229.\\
 E. P. Miles. Generalized Fibonacci Numbers and
associated Matrices, Amer.Math.Montly 67 (1960), 745-747.
\item[[21]]M.Mouline and M.Rachidi, Suites de Fibonacci generalisées et Chaines de
Markov,\\ Revista de la Real Academia de Ciencias Exatos, Fisicas
y Naturales de Madrid 89.1-2 (1995) 61-77.
\item[[22]]F.Dubeau, W.Motta, M.Rachidi and O.Sachi, On Weighted
r-GFS, \\The Fibonacci Quartely 35.2 (1997)102-110.\\ C.E Chidume,
M.Rachidi and E.H Zerouali, Solving the general truncated Moment
Problem by r-GFS method.\\ Journal of Mathematical Analysis and
Applications \bf{256}(2001)625-635.

\end{enumerate}

\end{document}